\documentclass[a4paper,12pt]{article}
\makeatletter

\usepackage[english]{babel}
\usepackage{lmodern}
\usepackage{microtype}

\usepackage{mathtools}
\usepackage{fullpage, euscript, amsfonts, amssymb, amsmath}
\usepackage{diffcoeff}
\usepackage[cal=boondox]{mathalfa}

\usepackage{setspace}
\usepackage[paper=a4paper, 
            inner=22mm, 
            outer=22mm, 
            top=22mm, 
            bottom=30mm, 
            headheight=6mm,
            includehead,
            includefoot, 
            heightrounded, 
            twoside]{geometry}
            
\usepackage{pdflscape}
\usepackage{placeins}
\usepackage{ragged2e}

\usepackage{graphicx}
\usepackage{float}

\usepackage{tabularx}
\usepackage{booktabs}
\usepackage{threeparttable}
\newcommand{\specialcell}[2][c]{\begin{tabular}[#1]{@{}c@{}}#2\end{tabular}}

\usepackage[
    backend=biber,
    style=authoryear,
    sorting=nyt,
    url=false,
    dashed=false]{biblatex}
\usepackage{csquotes}
\DeclareFieldFormat[techreport, report, misc]{title}{\mkbibquote{#1}}
\DeclareFieldFormat[techreport, report]{type}{\mkbibemph{#1}}
\DeclareFieldFormat{eprint}{\texttt{#1}}

\DeclareNameAlias{sortname}{family-given}

\AtBeginBibliography{\defcounter{maxnames}{99}}

\DeclareBibliographyDriver{*}{%
  \usebibmacro{bibindex}%
  \usebibmacro{begentry}%
  \printnames{author}%
  \setunit{\addcomma\space}%
  \printfield{year}%
  \setunit{\addcomma\space}%
  \printfield{title}%
  \newunit\newblock
  \printfield{journaltitle}%
  \setunit{\addcomma\space}%
  \printfield{volume}%
  \setunit{\addcomma\space}%
  \printfield{number}%
  \setunit{\addcomma\space}%
  \printfield{pages}%
  \newunit\newblock
  \printfield{institution}%
  \setunit{\addcomma\space}%
  \printfield{type}%
  \setunit{\addcomma\space}%
  \printfield{number}%
  \usebibmacro{finentry}%
}

\DeclareBibliographyDriver{article}{%
  \usebibmacro{bibindex}%
  \usebibmacro{begentry}%
  \printnames{author}%
  \setunit{\addspace}%
  \printtext[parens]{\printfield{year}}%
  \setunit{\addperiod\space}%
  \printfield{title}%
  \newunit\newblock
  \printfield{journaltitle}%
  \setunit{\addcomma\space}%
  \printfield{volume}%
  \setunit{\addcomma\space}%
  \printfield{number}%
  \setunit{\addcomma\space}%
  \printfield{pages}%
  \usebibmacro{finentry}%
}

\usepackage{multirow}
\usepackage{xcolor}
\definecolor{my_blue}{HTML}{1f77b4}

\usepackage{appendix}

\usepackage{bigfoot}

\usepackage[
    pdftitle={},
    pdfauthor={Hanfeng Chen, Maria Elena Filippin},
    hidelinks,
    hyperfootnotes=false
]{hyperref}
\hypersetup{
  colorlinks=false,
  citecolor=black,
  linkcolor=black}

\newcommand\blfootnote[1]{%
  \begingroup
  \renewcommand\thefootnote{}\footnote{#1}%
  \addtocounter{footnote}{-1}%
  \endgroup
}

\makeatother
\title{Central Bank Digital Currency: Demand Shocks and Optimal Monetary Policy}

\DeclareNewFootnote{AAffil}[arabic]
\DeclareNewFootnote{ANote}[fnsymbol]

\usepackage{etoolbox}
\makeatletter
\patchcmd\maketitle{\def\@makefnmark{\rlap{\@textsuperscript{\normalfont\@thefnmark}}}}{}{}{}

\makeatletter

\def\thanksAAffil#1{
  \footnotemarkAAffil\protected@xdef\@thanks{\@thanks%
        \protect\footnotetextAAffil[\the \c@footnoteAAffil]{#1}}%
}

\def\thanksANote#1{%
  \footnotemarkANote%
  \protected@xdef\@thanks{\@thanks%
        \protect\footnotetextANote[\the \c@footnoteANote]{#1}}%
}

\makeatother

\author{%
  Hanfeng Chen%
  \thanksAAffil{Uppsala University, National Institute of Economic Research, and Center for Monetary Policy and Financial Stability (CeMoF).}%
  $^{\,,\,}$\thanksANote{\textit{E-mails}: \texttt{hanfeng.chen@nek.uu.se} (H. Chen), \texttt{mariaelena.filippin@nek.uu.se} (M. E. Filippin)}%
  \and %
  Maria Elena Filippin%
  \thanksAAffil{Uppsala University, Central Bank of Ireland, and Center for Monetary Policy and Financial Stability (CeMoF).}%
  $^{\,,\,}$\footnotemarkANote[1]%
}

\date{October 13, 2025}

\begin{document}

\newgeometry{top=0.95in, bottom=0.95in, left=22mm, right=22mm}
\maketitle
\thispagestyle{empty}
\maketitle

\begin{abstract}
We study the implications of a central bank digital currency (CBDC) for the transmission of household preference shocks and for welfare in a New Keynesian framework where the CBDC competes with bank deposits for household resources and banks have market power. We show that an increase in the perceived benefit of CBDC has a mildly expansionary effect, weakening bank market power and significantly reducing the deposit spread. As households economize on liquid asset holdings, they reduce both CBDC and deposit balances. However, the degree of bank disintermediation is low, as deposit outflows remain modest. We then examine the welfare implications of CBDC rate setting and find that, compared to a non-interest-bearing CBDC, the gains with standard coefficients for a CBDC interest rate Taylor rule are modest, but they become considerable when the coefficients are optimized. Welfare gains increase with the CBDC benefit, and the optimal policy responses vary with the banking market structure.
\vspace{0.5cm}
\blfootnote{We are grateful to Mikael Carlsson, Daria Finocchiaro, Ulf Söderström, and Karl Walentin for their invaluable guidance and continuous support. For useful feedback and comments, we also thank Dirk Niepelt (discussant) and Anna Rogantini Picco (discussant), as well as conference and seminar participants at the 19th South-Eastern European Economic Research Workshop at the Bank of Albania and at the 2026 International Fintech Research Conference at the University of Pavia. This paper received the Best Paper Award at the 2026 International Fintech Research Conference. A large part of this work was conducted during our PhD at Uppsala University. The views expressed here are solely our own and do not necessarily reflect the views of the Central Bank of Ireland or the National Institute of Economic Research. 
}
\end{abstract}
\clearpage
\setcounter{page}{1}
\restoregeometry

\section{Introduction}

In recent years, central banks worldwide have begun to assess the potential benefits and risks of introducing central bank digital currencies (CBDCs). The idea of digital central bank money accessible to the wider public dates back to at least James Tobin in the 1980s [\textcite{Tobin1987}]. Since then, technological advancements have not only expanded the possibilities for CBDC but also heightened the demand for a public alternative to private digital money [\textcite{BIS2023}]. While early CBDC assessments and pilot projects have primarily taken place in smaller and developing countries, central banks in major economies have also launched extensive research and policy initiatives to explore the implementation of CBDCs [\textcite{ECB2024}].

The introduction of a CBDC raises important questions about household adoption and the perceived usefulness of digital central bank money. Given the lack of practical experience with CBDC in advanced economies, evidence on household preferences for CBDC remains limited. Two key aspects stand out: household perception of the benefits of CBDC, and their willingness to substitute between the CBDC and bank deposits. Rapid and unexpected shifts in these preferences could introduce new sources of macroeconomic fluctuations that complicate the policymakers' stabilization efforts.

One major concern in the CBDC debate is its interaction with the banking sector. Much of the discussion on a CBDC accessible to the wider public (i.e., retail) has focused on its potential competition with commercial bank deposits as a means of payment [see, e.g., \textcite{Assenmacher2021}, \textcite{Burlon2022}, \textcite{Brunnermeier2019}, \textcite{Chen2024}, \textcite{Chiu2023}, \textcite{Whited2023}, and \textcite{Williamson2022}]. Concerns primarily revolve around how competition from CBDC could affect bank lending, financial stability, and overall economic activity. Introducing a retail CBDC may disrupt credit intermediation, as deposits traditionally constitute a stable and significant source of bank funding [\textcite{Andolfatto2021}].

A CBDC may also have important implications for monetary policy, as a widely adopted digital central bank money could represent an additional instrument in the central bank toolkit. For instance, by adjusting the interest rate on CBDC, policymakers might directly influence household consumption and savings decisions. Additionally, the quantity of CBDC in circulation could affect financial intermediation [\textcite{Burlon2022}]. Thus, by adjusting the supply of CBDC, the central bank could gain an additional lever to affect the real economy. However, many central banks are currently considering issuing a non-interest-bearing CBDC to minimize risks to financial stability, making it as cash-like as possible [see, e.g., \textcite{CEPR2024}]. While such a design would limit the CBDC’s role as a monetary policy tool, its macroeconomic consequences remain uncertain. Understanding how different CBDC designs interact with bank market power, deposit pricing, and monetary policy transmission is therefore crucial.

Against this background, this paper addresses two key questions: (i) how do shocks to household preferences for CBDC transmit to the broader economy?, and (ii) how does CBDC design and its associated policy tools influence welfare? To address these questions, we extend a standard New Keynesian model by introducing a CBDC that competes with bank deposits. Households derive liquidity services from a combination of CBDC and deposits, making portfolio choices between the two. Consequently, liquid assets compete for household resources. Importantly, we model banks as non-competitive regional monopolists with pricing power in the deposit market. They issue deposits to finance investments in productive capital and reserves. We consider bank market power a key feature, in line with the growing literature emphasizing its importance for the transmission and efficacy of monetary policy [see, e.g., \textcite{Drechsler2017} and \textcite{Wang2022}].

First, we study the transmission of shocks to household preferences for CBDC, focusing on a shock to the CBDC benefit.\footnote{
In Appendix \ref{sec:app:eps}, we also examine a shock to the elasticity of substitution between the CBDC and deposits.
}
CBDC benefit captures a multitude of factors influencing the demand for CBDC: its liquidity benefits or ease of use as a means of payment, privacy considerations, payment security, or perceived riskiness. A positive shock can be interpreted as a flight to safety toward CBDC during stress periods, reflecting its safer nature as a central bank liability, or as an increase in the liquidity benefits that CBDC provides as a means of payment.

We find that a surprise $25\%$ increase in the CBDC benefit has mildly expansionary effects in our baseline scenario. Output, consumption, and inflation increase, while investment declines slightly. The deposit spread, representing the price of deposit products from the household's perspective, decreases significantly due to the combined effects of bank market power and CBDC competition. As the CBDC benefit increases, the household derives greater utility from it, weakening the bank's pricing power in the deposit market. Consequently, the deposit markup shrinks, leading to a decline in the deposit spread. Thus, greater CBDC benefit erodes bank market power. Interestingly, despite an increase in aggregate liquidity services, household holdings of CBDC and deposits decline, as the higher CBDC benefit allows households to economize on their liquid asset holdings. Importantly, the overall degree of bank disintermediation is low, as deposit outflows remain modest.

Second, we examine the welfare implications of CBDC interest rate policy. Traditionally, the central bank sets its standard monetary policy instrument, the nominal interest rate on government bonds, using a Taylor-type rule to stabilize inflation and output. In our framework, we assume that the central bank applies a similar Taylor-type rule to set the nominal interest rate on CBDC, and we investigate its welfare implications. To do so, we optimize Taylor-rule-based CBDC rate-setting under different banking structures and levels of CBDC benefit. For each specification, we first calculate the optimal Taylor-rule coefficients for the standard policy rate and its maximized conditional welfare, assuming a non-interest-bearing CBDC (i.e., gross CBDC rate fixed at $1$). We then assess the welfare differences between a non-interest-bearing CBDC and CBDC Taylor rules, using the optimal coefficients for the bond rate from the initial step. We follow the approach proposed by \textcite{Schmitt2007} and \textcite{Faia2007} for optimizing monetary policy rules, and numerically compute the Taylor rule coefficients for inflation and output that maximize the expected value of household welfare.

Our findings indicate that a non-optimized Taylor-rule-based CBDC (i.e., a CBDC Taylor rule with baseline coefficients) yields small welfare improvements relative to a non-interest-bearing CBDC, but the gains are considerable when the coefficients are optimized. Welfare gains are higher when the CBDC benefit is higher. The optimal policy response varies across banking structures. With monopolist banks, the optimized interest rate response to inflation depends on the benefit of CBDC, while the magnitude of the response to output is consistent across specifications. With competitive banks, we get the stark result that the CBDC rate should focus entirely on output. Lastly, our results suggest that interest rate smoothing for CBDC is not desirable.\\

\noindent
\textbf{Related literature}. \hspace{0.5mm} Our paper contributes to the literature on the relationship between CBDC and bank deposits by analyzing the transmission of shocks to household preferences for CBDC, similar to \textcite{Ferrari2022}, who incorporate a CBDC into a standard two-country dynamic stochastic general equilibrium (DSGE) model with financial frictions to examine its role in the international transmission of monetary policy and technology shocks. They find that the CBDC amplifies international spillovers of shocks, increasing global economic linkages. However, the extent of these effects is highly dependent on the specific design features of the CBDC. In another work, \textcite{Agur2022} consider CBDC, bank deposits, and cash as imperfect substitutes with horizontal differentiation of means of payment and find that the extent of bank disintermediation depends on how closely the CBDC competes with deposits [see also \textcite{Andolfatto2021}]. Similarly, \textcite{Keister2022} consider a competitive market and show that a deposit-like CBDC tends to crowd out bank deposits while increasing the aggregate stock of liquid assets in the economy, promoting more efficient levels of production and exchange, and ultimately raising welfare. In line with \textcite{Keister2022}, we also find that aggregate liquidity services increase following a shock to the CBDC benefit, but as households economize on liquid asset holdings, they reduce both CBDC and deposit balances. However, the degree of bank disintermediation is low, as deposit outflows remain modest.

There is no consensus on the effect of CBDC on banks: some studies conclude that CBDC does not necessarily have a negative impact on bank lending [see, e.g., \textcite{Andolfatto2021}; \textcite{Chiu2023}; and \textcite{Whited2023}], while others show that CBDC impacts on banks depend on its design [see, e.g., \textcite{Assenmacher2021}; \textcite{Burlon2022}; \textcite{Keister2022}; and \textcite{Williamson2022}]. Our results align more closely with the view that CBDC does not inherently cause bank disintermediation, as even in the case of a higher CBDC benefit, deposit outflows remain limited, mitigating potential disruptions to financial intermediation.

This paper also contributes to the literature examining the consequences of introducing a CBDC on monetary policy. For instance, using a New Keynesian model with heterogeneous banks and a frictional interbank market, \textcite{Abad2023} find that for moderate CBDC adoption, the reduction in bank deposits is absorbed almost one-to-one by a decline in central bank reserves, leading to a transition from a floor system with ample reserves to a corridor system with reduced reserves, without significantly affecting bank lending or GDP. However, higher levels of CBDC adoption may require the central bank to provide more credit to banks, leading to a ceiling system with scarce reserves. Similarly, \textcite{Bhattarai2024} introduce a CBDC which competes with bank deposits as the household's source of liquidity into a New Keynesian framework with financial frictions. They find that introducing a fixed-interest-rate CBDC magnifies the effects of monetary policy shocks on output and consumption by altering liquidity dynamics, but the specific impact depends on the type of monetary policy framework and the central bank’s balance sheet rules.

CBDC introduction may affect financial stability and the resilience of the banking system, which in turn influences the effectiveness of monetary policy. \textcite{Boser2024} show that while an interest-bearing CBDC can enhance monetary policy transmission and financial stability by offering a safer form of money, it could also pose risks, such as the potential for digital runs on banks as depositors shift from traditional deposits to CBDC for safety. These risks can initially be mitigated through tight collateral requirements and default penalties, enhancing bankers' monitoring incentives and productivity. Nevertheless, widespread CBDC adoption can increase liquidity risks, potentially destabilizing traditional banking and reducing the long-term effectiveness of CBDC in improving banking practices and economic welfare. Developing a general equilibrium model of payments and banking, \textcite{Chiu2023b} show that a cash-like CBDC is more effective than a deposit-like CBDC in promoting consumption and welfare while increasing bank intermediation through improved payment efficiency. By introducing a feedback mechanism between transactions and deposit creation in their model, they demonstrate that a well-designed CBDC can crowd in banking even without market power among banks. In another study, \textcite{Davoodalhosseini2022} uses a New Monetarist framework to compare scenarios where only cash, only CBDC, or both are available. His findings suggest that CBDC can lead to more efficient economic allocations than cash if its usage costs are not prohibitive. However, having cash and CBDC available may result in lower welfare than in scenarios where only one form of currency is used.

Our model is similar to \textcite{Bhattarai2024}, as we also employ a New Keynesian model where CBDC and deposits are substitutable, but we consider non-competitive banks to capture the role of bank market power in monetary transmission when CBDC is used as a policy tool. Our findings show that optimizing CBDC interest rate policies through a Taylor rule improves household welfare relative to a non-interest-bearing CBDC, particularly when the CBDC is perceived as highly beneficial, and that the optimal CBDC interest rate policy response varies across banking structure. \\

The rest of the paper is organized as follows. Section \ref{s:model} introduces the model. Section \ref{s:calib} presents the baseline calibration. Section \ref{s:IRFs} examines the transmission of household preference shocks. Section \ref{s:opt_mp} explores the welfare implications of CBDC interest rate policies. Section \ref{s:the_end_and_the_death} concludes.

\section{A New-Keynesian model with CBDC and deposits}\label{s:model}
We extend a standard New-Keynesian model by introducing CBDC and bank deposits. Households value a combination of CBDC and deposits for liquidity purposes, invest in government bonds, and purchase physical capital from capital producers. Banks issue deposits to fund investments in physical capital and reserves. Following \textcite{Niepelt2024}, a role for reserves is introduced by assuming that higher holdings of reserves relative to deposits reduce banks' cost of issuing debt. Banks are regional monopolists in their local deposit markets and thus have pricing power over deposit products. Intermediate good firms use capital and labor as inputs to produce intermediate goods, and rent capital from households and banks. Retail firms costlessly differentiate intermediate goods and sell the differentiated goods to final good firms. Retail firms are subject to price rigidity following \textcite{Calvo1983} and \textcite{Yun1996}. The consolidated government determines the interest rate on government bonds and CBDC and supplies CBDC elastically to meet demand.

\subsection{Households}
The economy is populated by a continuum of identical and infinitely lived households of unit mass. The household preferences are defined over consumption, $c_t$, liquidity services, $z_{t+1}$, and hours worked, $l_t$. Preferences are given by the utility function  
\begin{align}
    \mathcal{U}(c_t, z_{t+1}, l_t)=\frac{\left((1-v)c_t^{1-\psi}+vz_{t+1}^{1-\psi}\right)^{\frac{1-\sigma}{1-\psi}}-1}{1-\sigma} - \xi\frac{l_t^{1+\iota}}{1+\iota}, 
\end{align}
where $v\in (0,1)$ is the relative utility weight on liquidity; $\sigma>0$ is the inverse elasticity of substitution between consumption-liquidity bundles at different dates; $\psi>0$ is the inverse intratemporal elasticity of substitution between consumption and liquidity; $\xi>0$ is the disutility of labor; and $\iota>0$ is the inverse Frisch elasticity.\footnote{New-Keynesian models are typically derived assuming separability between consumption and liquidity, i.e., $\sigma=\psi$. Instead, we develop the model for general non-separability because typical estimates of liquidity demand, $1/\psi$, are lower than the standard values for the intertemporal elasticity of substitution, $1/\sigma$. In other words, the separability assumption may be overly restrictive [\textcite{Piazzesi2022_money}].}

We assume that the household derives liquidity services from real holdings of CBDC, $m_{t+1}$, and bank deposits, $n_{t+1}$. Liquidity is aggregated according to a constant elasticity of substitution aggregator:
\begin{align}
    z_{t+1}(m_{t+1},n_{t+1}) = \left( \lambda_t m^{1-\epsilon_t}_{t+1} + n^{1-\epsilon_t}_{t+1} \right)^\frac{1}{1-\epsilon_t}, 
\end{align}
where $\lambda_t \geq 0$ indexes the household's preferences for CBDC relative to deposits and $\epsilon_t\geq 0$ is the inverse elasticity of substitution between the CBDC and deposits. The two parameters are allowed to be time-varying, as indicated by their time subscripts. The relative preference for CBDC, $\lambda_t$ (representing the CBDC benefit), captures a multitude of factors affecting the demand for CBDC, such as its liquidity benefits or ease of use as a means of payment, privacy considerations, payment security, or perceived riskiness. 

In addition to CBDC and deposits, the household can invest in government bonds, $b_{t+1}$, and purchase physical capital, $k_{t+1}^h$, at unit price $q_{t}$, from specialized capital producers. At the end of the period, the household sells the stock of undepreciated capital back to capital producers. The period budget constraint of the household, in real terms, is given by 
\begin{align}
    c_t + m_{t+1} + n_{t+1} + b_{t+1} + q_{t}k_{t+1}^h +\tau_t &= w_tl_t + d_t + \frac{m_t R^m_t}{\pi_t} + \frac{n_t R^n_t}{\pi_t}+\frac{b_t R_t}{\pi_t} \nonumber\\
    & \hspace{0.5cm} +q_{t-1}k_t^h R^k_t + (q_t-\delta)k_t^h, 
\end{align}
where $\tau_t$ is the lump-sum tax net of government transfers; $w_t$ is the wage rate; $d_t$ is the distributed profit from banks and firms owned by the household; $R^m_t$, $R^n_t$ and $R_t$ are the nominal gross interest rates on CBDC, deposits and bonds, respectively; $R_t^k$ is the return on capital; $\pi_t=p_t/p_{t-1}$ is the gross rate of inflation; and $\delta$ is the rate of capital depreciation. We assume the nominal rates on CBDC, deposits, and bonds are risk-free, while the return on capital may be risky. The household maximizes the discounted sum of lifetime utility 
\begin{align}
    \mathbb{E}_t \sum_{j=0}^\infty \beta^j \mathcal{U}(c_{t+j},z_{t+1+j},l_{t+j}),
\end{align}
by choosing consumption, hours worked, and assets, subject to the budget constraint. 

We define the interest rate spread on liquid asset $i$ as
\begin{align}
    \chi^i_{t+1} = \frac{R_{t+1}-R^i_{t+1}}{R_{t+1}}, \qquad i \in \{m, n\}.
\end{align}
The interest spreads constitute the household's opportunity costs of holding CBDC and deposits since they show the discounted foregone interest associated with each asset. The first-order conditions with respect to CBDC and deposits imply a demand for aggregate liquidity given by 
\begin{align}
     z_{t+1} &= c_{t}\left(\frac{v}{1-v}\frac{1}{\chi_{t+1}^z}\right)^{\frac{1}{\psi}}, 
\end{align}
where we denote $\chi_{t+1}^z$ as the average cost of liquidity:
\begin{align}
    \chi_{t+1}^z= \frac{\chi_{t+1}^m\chi_{t+1}^n}{\left(\lambda_t^{\frac{1}{\epsilon_t}}\left(\chi_{t+1}^n\right)^{\frac{1-\epsilon_t}{\epsilon_t}}+\left(\chi_{t+1}^m\right)^{\frac{1-\epsilon_t}{\epsilon_t}}\right)^{\frac{\epsilon_t}{1-\epsilon_t}}}.  \label{avgcl}
\end{align}
Given the demand for aggregate liquidity, the demand for CBDC is
\begin{align}
    m_{t+1} = z_{t+1}\left(\lambda_t\frac{\chi_{t+1}^z}{\chi_{t+1}^m}\right)^{\frac{1}{\epsilon_t}}, \label{m}
\end{align}
and the demand for deposits is
\begin{align}
    n_{t+1} = z_{t+1}\left(\frac{\chi_{t+1}^z}{\chi_{t+1}^n}\right)^{\frac{1}{\epsilon_t}}. \label{n}
\end{align}
Using (\ref{m}) and (\ref{n}), we can also express the average cost of liquidity, equation (\ref{avgcl}), as
\begin{align}
    \chi_{t+1}^z= \frac{m_{t+1}}{z_{t+1}}\chi_{t+1}^m+\frac{n_{t+1}}{z_{t+1}}\chi_{t+1}^n.
\end{align}
In other words, $\chi_{t+1}^z$ is a weighted average where the weights are the relative shares of each asset. Combining equations (\ref{m}) and (\ref{n}) yields an expression for relative demand for CBDC:
\begin{align}
    \frac{m_{t+1}}{n_{t+1}} = \left(\lambda_t\frac{\chi_{t+1}^n}{\chi_{t+1}^m}\right)^{\frac{1}{\epsilon_t}}. \label{mn}
\end{align}
We see that the household demands more CBDC relative to deposits if, everything else equal, the relative price of CBDC, $\chi_{t+1}^m/\chi_{t+1}^n$, is lower. Moreover, the relative demand for CBDC is increasing in its benefit, $\lambda_t$. The extent to which the elasticity of substitution between the two assets, $1/\epsilon_t$, affects relative demand depends on the sizes of the interest spreads, $\chi_{t+1}^m$ and $\chi_{t+1}^n$. For example, suppose the deposit spread is larger than the CBDC spread. Then, a higher elasticity of substitution would imply a higher ratio of CBDC to deposits. 

The household's Euler equations for consumption and capital are given by, respectively,  
\begin{align}
    \mathcal{U}_{c,t} &= \beta \mathbb{E}_t \mathcal{U}_{c,t+1}\frac{R_{t+1}}{\pi_{t+1}}, \label{eecons} \\
    \mathcal{U}_{c,t} &= \beta \mathbb{E}_t \mathcal{U}_{c,t+1}\left(R_{t+1}^k+\frac{q_{t+1}-\delta}{q_t}\right). 
\end{align}
In equation (\ref{eecons}), $\mathcal{U}_{c,t}$ is the time-$t$ marginal utility of consumption:
\begin{align}
    \mathcal{U}_{c,t} = c_t^{-\sigma}\Omega_t,
\end{align}
which depends on the average cost of liquidity captured by the term $\Omega_t$: 
\begin{align}
    \Omega_t &= (1-v)^{\frac{1-\sigma}{1-\psi}}\left(1+\left(\frac{v}{1-v}\right)^{\frac{1}{\psi}}\left(\chi_{t+1}^z\right)^{1-\frac{1}{\psi}}\right)^{\frac{\psi-\sigma}{1-\psi}}. 
\end{align} 
Relative to a standard specification without liquidity preference, the term $\Omega_t$ captures the impact of liquidity on the marginal utility of consumption. $\Omega_t$ can also be interpreted as an ideal price index for a bundle of consumption and liquidity, and it depends on the average cost of liquidity and, in turn, on the interest spreads. Lastly, the labor supply condition reads: 
\begin{align}
    -\mathcal{U}_{l,t} &= w_t \mathcal{U}_{c,t}, 
\end{align}
where $\mathcal{U}_{l,t}$ is the disutility of supplying labor 
\begin{align}
    \mathcal{U}_{l,t} = -\xi l_t^\iota. 
\end{align}

\subsection{Banks}
Banks invest in capital, $k_{t+1}^b$, and reserves, $r_{t+1}$, and fund themselves using deposits, $n_{t+1}$, and equity, $e_b$. The balance sheet of a bank is, in real terms:  
\begin{align}
    q_{t}k_{t+1}^b + r_{t+1} = n_{t+1} + e_b. \label{bank_bs}
\end{align}
In this setting, we assume each bank is endowed with a fixed amount of equity so that balance sheet adjustments are made using only debt. We follow \textcite{Niepelt2024} in assuming a banking sector segmented by region, with one monopolist bank in each segment. There exists a finite number of regions of equal size. A household only holds deposits with its regional bank, but the regions do not otherwise restrict other types of economic transactions. Since banks are identical, we refer to the whole banking sector as a representative bank.

A role for reserves is introduced by assuming that the bank incurs a cost per unit of deposits issued. We use a simplified version of the cost function in \textcite{Niepelt2024} where the per unit cost, $\omega_t$, is decreasing in the bank's liquidity ratio, $\zeta_{t+1}$:
\begin{align}
    \omega_t = \phi\zeta_{t+1}^{1-\varphi}, \label{bank_cost}
\end{align}
where $\phi \geq 0$; $\varphi>1$; and $\zeta_{t+1}=r_{t+1}/n_{t+1}$. When a bank has a higher stock of deposits, it requires more liquid assets (reserves) to settle payments on behalf of its depositors. If these reserves are insufficient, the bank must use less liquid assets (such as capital) for settlements, potentially leading to losses. Alternatively, this can be seen more generally as a simple way to model the increasing cost of issuing debt [\textcite{Piazzesi2022_money}].   

The bank faces the same rate of return on capital as the household and sells the undepreciated capital back to capital producers at the end of each period. The real profit of the bank is given by 
\begin{align}
    d_t^b = q_{t-1}k_t^bR_t^k + r_t\frac{R_t^r}{\pi_t} + (q_t-\delta)k_t^b-n_t\frac{R_t^n}{\pi_t}-e_b-n_{t+1}\omega_t, 
\end{align}
where $R_t^r$ is the nominal gross rate of interest on reserves. The bank is infinitely lived and maximizes the discounted sum of profits
\begin{align}
    \mathbb{E}_t \sum_{j=0}^\infty \Lambda_{t,t+j} d_{t+j}^b, 
\end{align}
subject to the demand for its deposits, given by equation (\ref{n}), its balance sheet, equation (\ref{bank_bs}), and the cost function, equation (\ref{bank_cost}). The bank chooses the quantities of capital, reserves, and deposits and discounts the profit with the household real stochastic discount factor, $\Lambda_{t,t+j}=\beta^j\mathcal{U}_{c,t+j}/\mathcal{U}_{c,t}$.

The first-order condition with respect to reserves gives the bank's optimal liquidity ratio:
\begin{align}
    \zeta_{t+1} = \left(\frac{\chi_{t+1}^r}{\phi(\varphi-1)}\right)^{-\frac{1}{\varphi}}, \label{zeta2} 
\end{align}
where $\chi_{t+1}^r=1-R_{t+1}^r/R_{t+1}$ is the interest spread on reserves. The reserve spread has a similar interpretation to the CBDC and deposit spreads for the household. It shows the discounted forgone interest that the bank incurs when holding reserves and thus represents the opportunity cost of reserves. Note that the bank's desired liquidity ratio decreases in the reserve spread. As the cost of reserves increases, the bank reduces its liquidity ratio, and consequently, the unit cost of issuing deposits increases. The first-order condition with respect to deposits yields a pricing equation that determines the deposit spread 
\begin{align}
    \chi_{t+1}^n-\chi_{t+1}^n\left(\frac{1-s_t}{\psi}+\frac{s_t}{\epsilon_t}\right)^{-1} = \varphi\phi\zeta_{t+1}^{1-\varphi}, \label{xn}
\end{align}
where $s_t \in (0,1)$ is given by
\begin{align}
    s_t = \lambda_t^{\frac{1}{\epsilon_t}}\left(\frac{\chi_{t+1}^z}{\chi_{t+1}^m}\right)^{\frac{1-\epsilon_t}{\epsilon_t}}. \label{s}
\end{align}
Equation (\ref{xn}) has the classic interpretation that the price of deposits, given by the deposit spread, depends on the marginal cost of issuing deposits and a markup term. The right-hand side of the equation (\ref{xn}) shows the bank's cost of issuing debt. The cost of debt is decreasing in the liquidity ratio, $\zeta_{t+1}$ (recall that $\varphi>1$). A greater quantity of reserves relative to that of deposits reduces the bank's cost of providing liquidity to the household and thus decreases the deposit spread. As described earlier, a higher reserve spread would increase the cost of deposit issuance since it raises the bank's opportunity cost of reserves and incentivizes the bank to reduce the liquidity ratio. Since the bank is a monopolist, it can charge a markup over the marginal cost. The second term on the left-hand side of (\ref{xn}) constitutes the markup over the marginal cost, which is inversely related to the elasticity of demand for deposits given by 
\begin{align}
    -\left(\frac{1-s_t}{\psi}+\frac{s_t}{\epsilon_t}\right). \label{bank_elas}
\end{align}
Equation (\ref{bank_elas}) shows that the sensitivity of deposit demand to the deposit spread depends on two factors: the elasticity of substitution between consumption and deposits, $1/\psi$, and the elasticity of substitution between CBDC and deposits, $1/\epsilon_t$. The relative importance of the two, in turn, depends on how expensive CBDC is relative to the average cost of liquidity captured by $s_t$. A higher CBDC spread, for example, means that the substitutability between CBDC and deposits now matters more for the bank's ability to charge a markup. 

\subsection{Firms}
\subsubsection{Final good firms}
Final good firms are competitive and use differentiated goods, $y_t^i$, to produce the final output, $y_t$. The production technology of a final good firm is 
\begin{align}
    y_t = \left(\int_0^1 \left(y_t^i\right)^{1-\eta_t} di \right)^{\frac{1}{1-\eta_t}}, \label{agg_y}
\end{align}
and $\eta_t$ is the inverse elasticity of substitution across good varieties. We let $\eta_t$ be time-varying as a simple way to allow for cost-push shocks in the model. The real profit of a final good firm is 
\begin{align}
    d_t^f = y_t - \int_0^1 \frac{p_t^iy_t^i}{p_t} di, 
\end{align}
where $p_t$ is the price of the final output and $p_t^i$ is the price of differentiated goods indexed by $i$. The firm maximizes its profit by choosing its demand for each good, subject to the technology given by equation (\ref{agg_y}). The first-order conditions of the final good firm deliver the demand for differentiated goods
\begin{align}
    y_t^i = y_t\left(\frac{p_t^i}{p_t}\right)^{-\frac{1}{\eta_t}}, \label{y_i}
\end{align}
where $p_t$ can be shown to be a price index given by 
\begin{align}
    p_t = \left(\int_0^1 \left(p_t^i\right)^{\frac{\eta_t-1}{\eta_t}} di\right)^{\frac{\eta_t}{\eta_t-1}}. \label{p_ind}
\end{align}

\subsubsection{Intermediate good firms}
Competitive intermediate good firms produce a common intermediate output, $y_t^m$, using a constant return to scale technology, with capital and labor as inputs. The production function is given by 
\begin{align}
    y_t^m = a_tk_t^\alpha l_t^{1-\alpha}, 
\end{align}
where $k_t$ is the firm's current capital stock; $l_t$ is demanded labor; $a_t$ is productivity; and $\alpha$ is the capital share of output. 
The intermediate goods are eventually sold to retail firms at the price $p_t^m$. The real profit of the firm is:
\begin{align}
    d_t^m = \frac{y_t^mp_t^m}{p_t} - q_{t-1}k_tR_t^k - w_tl_t.  
\end{align}
Each period, the firm chooses the current labor demand and the next-period capital stock to maximize its discounted profits
\begin{align}
    \mathbb{E}_t \sum_{j=0}^\infty \Lambda_{t,t+j} d_{t+j}^m. 
\end{align}
The first-order condition with respect to labor determines the firm's labor demand as
\begin{align}
    w_t = a_t(1-\alpha)\gamma_t\left(\frac{k_t}{l_t}\right)^\alpha, 
\end{align}
where $\gamma_t=p_t^m/p_t$. The first-order condition with respect to capital yields capital demand as
\begin{align}
    R_t^k = \frac{a_t\alpha \gamma_t(k_t/l_t)^{\alpha-1}}{q_{t-1}}. 
\end{align}

\subsubsection{Retail firms}
Each good variety $i\in [0,1]$ is produced by a monopolistically competitive retail firm. The retail firm buys intermediate goods and costlessly differentiates them into its variety, i.e., $y_t^i=y_t^m$. 

The real profit of a retail firm is the difference between the revenues from selling its good and the cost of buying the intermediate good:
\begin{align}
    d_t^{r,i} = \frac{p_t^iy_t^i}{p_t} - \gamma_ty_t^m,   
\end{align}
where $\gamma_t$ is now interpreted as the real marginal cost of the retail firm. Nominal rigidity is introduced by ways of \textcite{Calvo1983} and \textcite{Yun1996}. Specifically, in each period, retail firms face a probability $\theta\in [0,1)$ of not being able to change their price, i.e., $p_t^i = p^i_{t-1}$. A retail firm maximizes the discounted sum of profits by choosing its price, taking into account the probability of not being able to re-optimize: 
\begin{align}
    \mathbb{E}_t \sum_{j=0}^\infty \theta^j \Lambda_{t,t+j}d_{t+j}^{r,i}, 
\end{align}
subject to the demand schedule for its good, given by equation (\ref{y_i}). 

The first-order condition of the retail firm is 
\begin{align}
    p_t^i = \frac{1}{1-\eta_t}\frac{\mathbb{E}_t\sum_{h=0}^\infty(\theta\beta)^h\mathcal{U}_{c,t+h} y_{t+h} p_{t+h}^{\frac{1}{\eta_t}}\gamma_{t+h}}{\mathbb{E}_t\sum_{h=0}^\infty(\theta\beta)^h\mathcal{U}_{c,t+h} y_{t+h} p_{t+h}^{\frac{1}{\eta_t}-1}}. 
\end{align}
The latter equation shows that every retail firm will choose the same reset price. Let $p_t^r=p_t^i$ denote the common reset price and $\pi_t^r = p_t^r/p_{t-1}$ be the reset inflation. Rewriting the retail firm's first-order condition more compactly in terms of inflation:
\begin{align}
    \pi_t^r = \frac{1}{1-\eta_t}\pi_t\frac{x_{1,t}}{x_{2,t}}, 
\end{align}
where 
\begin{align}
    x_{1,t} &= \mathcal{U}_{c,t} y_t \gamma_t+\theta\beta \mathbb{E}_t\pi_{t+1}^{\frac{1}{\eta_t}}x_{1,t+1}, \\
    x_{2,t} &= \mathcal{U}_{c,t} y_t + \theta\beta \mathbb{E}_t \pi_{t+1}^{\frac{1}{\eta_t}-1}x_{2,t+1}. 
\end{align}

\subsubsection{Capital producers}
Competitive capital producers buy undepreciated capital at the end of each period, make capital investment decisions, and sell newly built capital. Let $I_t^n$ be the net capital investment. The gross investment, $I_t$, is given by
\begin{align}
    I_t = I_t^n+\delta k_t. 
\end{align}
Note that we assume depreciated capital is refurbished in its entirety every period. The real profit of a capital producer in terms of net investments is
\begin{align}
    d_t^c = (q_t-1)I_t^n - \frac{\theta_c}{2}\left(\ln\left( \frac{I_t^n+\bar{I}}{I_{t-1}^n+\bar{I}} \right)\right)^2 \left(I_t^n+\bar{I}\right), 
\end{align}
where $\theta_c\geq 0$ controls the adjustment cost associated with investments and $\bar{I}$ is the steady-state level of gross investment. A capital producer maximizes the discounted profits by choosing the level of net investment:
\begin{align}
    \mathbb{E}_t \sum_{j=0}^\infty \Lambda_{t,t+j}d_{t+j}^{c}. 
\end{align}
The first-order condition of the capital producer yields an expression for the price of capital:
\begin{align}
    q_t = 1+\frac{\theta_c}{2}\left(\ln\left(\frac{I_{t}^n+\bar{I}}{I_{t-1}^n+\bar{I}}\right)\right)^2 +\theta_c\ln\left(\frac{I_t^n+\bar{I}}{I_{t-1}^n+\bar{I}}\right)-\mathbb{E}_t\Lambda_{t,t+1}\theta_c\ln\left(\frac{I_{t+1}^n+\bar{I}}{I_t^n+\bar{I}}\right)\frac{I_{t+1}^n+\bar{I}}{I_t^n+\bar{I}}.
\end{align}

\subsection{Consolidated government}\label{govt}
The government issues bonds, reserves, and CBDC to finance its primary deficit, interest payments on existing liabilities, and costs of issuing (and managing) CBDC and reserves. Following \textcite{Niepelt2024}, we assume that the government incurs a per-unit cost of issuing CBDC, $\mu_m$, and a per-unit cost of issuing reserves, $\mu_r$. Its budget constraint is given by 
\begin{align}
    b_{t+1}+r_{t+1}+m_{t+1} = g_t-\tau_t + \frac{b_tR_t}{\pi_t}+\frac{r_tR_t^r}{\pi_t}+\frac{m_tR_t^m}{\pi_t} + r_{t+1}\mu_r + m_{t+1}\mu_m, 
\end{align}
where $g_t$ denotes government spending. The government sets the nominal interest rate on bonds, $R_{t+1}$, as its standard monetary policy instrument. The bond rate follows a Taylor-type rule that seeks to stabilize inflation and output:
\begin{align}
    \ln(R_{t+1}) = (1-\rho_R)\ln(\bar{R}) + \rho_R\ln(R_{t}) + (1-\rho_R)\left(\theta_{\pi} \ln\left(\frac{\pi_t}{\bar{\pi}}\right) + \theta_y\ln\left(\frac{y_t}{\bar{y}}\right) \right)+e_t^R, \label{r_taylor}
\end{align}
where $\bar{R}$ is the steady-state bond rate; $\bar{\pi}$ is the inflation target; $\bar{y}$ is the steady-state output; $\rho_R \in [0,1]$ captures monetary policy inertia; $\theta_{\pi}\geq 0$ and $\theta_{y}\geq 0$ are responses to inflation and output fluctuations, respectively; and $e_t^R$ is an exogenous one-time monetary policy innovation. We assume the government sets the reserve rate such that there is a constant spread between it and the policy rate:
\begin{align}
    R_{t+1}^r = R_{t+1}(1-\bar{\chi}^r), \label{Rr}
\end{align}
where $\bar{\chi}^r$ denotes the steady-state reserve spread.

Given the focus of our paper, we assume that the government sets the nominal interest rate on CBDC and supplies it elastically to meet demand. The government sets the CBDC rate according to a Taylor-type rule:
\begin{align}
    \ln(R_{t+1}^m) = (1-\rho_R^m)\ln(\bar{R}^m) + \rho_R^m\ln(R_{t}^m) + (1-\rho_R^m)\left(\theta_{\pi}^m \ln\left(\frac{\pi_t}{\bar{\pi}}\right) + \theta_y^m\ln\left(\frac{y_t}{\bar{y}}\right) \right)+e_t^{m}, \label{rm_taylor}
\end{align}
where $\bar{R}^m$ is the steady-state CBDC rate; $\rho_R^m \in [0,1]$ is the CBDC rate smoothing parameter; $\theta_{\pi}^m\geq 0$ and $\theta_{y}^m\geq 0$ are the CBDC rate responses to inflation and output fluctuations, respectively; and $e_t^{m}$ is the one-time CBDC exogenous innovation. 

\subsection{Market clearing and aggregation}
Capital market clearing requires the total stock of capital to be equal to the sum of household and bank capital holdings:
\begin{align}
    k_t = k_t^h+k_t^b. 
\end{align}
Aggregating over all differentiated goods gives 
\begin{align}
    \int_0^1 y_t^i di = \int_0^1 y_t^m di = \int_0^1 y_t\left(\frac{p_t^i}{p_t}\right)^{-\frac{1}{\eta_t}} di,  
\end{align}
from which we find an expression for aggregate output $y_t$
\begin{align}
    y_t = \frac{a_tk_t^\alpha l_t^{1-\alpha}}{v_t^p},
\end{align}
and price dispersion
\begin{align}
    v_t^p = \int_0^1 \left(\frac{p_t^i}{p_t}\right)^{-\frac{1}{\eta_t}} di = (1-\theta)\left(\frac{\pi_t}{\pi_t^r}\right)^{\frac{1}{\eta_t}}+\pi_t^{\frac{1}{\eta_t}}\theta v_{t-1}^p. 
\end{align}
Using the aggregate price index (\ref{p_ind}) and the Calvo assumption we can show that 
\begin{align}
    \pi_t^{1-\frac{1}{\eta_t}} = (1-\theta)\left(\pi_t^r\right)^{1-\frac{1}{\eta_t}}+\theta. 
\end{align}

The total profit distributed to the household must be equal to the sum of profits from banks and firms:
\begin{align}
    d_t = d_t^b + d_t^f + d_t^m + \int_0^1 d_t^r di + d_t^c. 
\end{align}
Lastly, combining the budget constraints of the household and the government and the total profit, we derive the aggregate resource constraint as
\begin{align}
    c_t + g_t + I_t = y_t - \frac{\theta_c}{2}\left(\ln\left( \frac{I_t^n+\bar{I}}{I_{t-1}^n+\bar{I}} \right)\right)^2 \left(I_t^n+\bar{I}\right) - m_{t+1}\mu_m - n_{t+1}(\omega_t+\zeta_{t+1}\mu_r),
\end{align}
and the law of motion for capital 
\begin{align}
    k_{t+1}=k_t+I_t^n.
\end{align}

\subsection{Structural shocks}
One of the major sources of uncertainty for central banks when designing CBDC is how households would perceive it in relation to commercial bank deposits. Therefore, we allow the two key household preference parameters—the benefit of CBDC, $\lambda_t$, and the inverse elasticity of substitution between CBDC and deposits, $\epsilon_t$—to be time-varying and subject to shocks. In addition, we also consider shocks to productivity, $a_t$, government spending, $g_t$, and the inverse elasticity of substitution across good varieties, $\eta_t$, which is a simple way to allow for fluctuations in the desired markup of retail firms. They all follow a log AR(1) shock of the form
\begin{align}
     \ln(\upsilon_t) = (1-\rho_\upsilon)\ln(\bar{\upsilon}) + \rho_\upsilon \ln(\upsilon_{t-1}) + e_t^\upsilon, \qquad \upsilon \in \{\lambda, \epsilon, a, g, \eta \},
\end{align}
where $\bar{\upsilon}$ is the steady-state value; $\rho_\upsilon$ is the persistence parameter; and $e_t^\upsilon$ is the exogenous one-time innovation. Lastly, we also allow for shocks to conventional monetary policy, $e_t^R$, and CBDC rate, $e_t^{m}$, as detailed in Section \ref{govt}. 

\section{Calibration}\label{s:calib}
We calibrate the model to the U.S. economy and interpret each model period as a quarter. We use variables with an overbar to denote their steady-state values. Table \ref{cal_params} shows the chosen targets and the baseline calibration. 

\subsection{Households}
The household's discount factor, $\beta$, is set to the standard value of $0.99$. We set the inverse intertemporal elasticity of substitution, $\sigma$, to $1$, which is standard in the literature. Following \textcite{Piazzesi2022_money}, the inverse elasticity of substitution between consumption and liquidity, $\psi$ is calibrated to $4.55$. The inverse Frisch elasticity, $\iota$, is set to $1$, in line with the estimate of \textcite{Chetty2011}. We calibrate the utility weight of liquidity, $v$, to $0.08$ to match a steady-state quarterly liquidity-to-output ratio of $1.04$ [see, e.g., \textcite{bayer}]. The disutility of labor, $\xi$, is set to $9.46$ to target a steady-state labor of $1/3$.

For the CBDC parameters governing households' preferences, we do not have data. However, many central banks are considering proposals such that the CBDC payment transactions would be subject to Anti-Money Laundering and Combating the Financing of Terrorism (AML/CFT) requirements, similar to private digital means of payment\footnote{See the Frequently Asked Questions at \href{https://finance.ec.europa.eu/digital-finance/digital-euro/frequently-asked-questions-digital-euro-and-legal-tender-cash_en}{European Commission Digital Euro FAQs}.}, and the user experience would be largely the same for CBDC or deposit-based payments. Therefore, in the baseline, we assume the steady-state benefit of CBDC is $\bar{\lambda}=1$, implying that each CBDC unit provides the same benefit as deposits. We set the steady-state inverse elasticity of substitution between CBDC and deposits, $\bar{\epsilon}$, to $1/6$. This corresponds to a medium degree of substitutability following \textcite{Bacchetta2022}. The persistence parameters of the AR(1) shock describing the preference shocks, $\lambda_t$ and $\epsilon_t$, are assumed to be $0.9$. 

\subsection{Banks and firms}
The first parameter in the bank's cost function, $\phi$, is calibrated to $0.0008$ to match the liquidity ratio, $\zeta=0.1945$, estimated by \textcite{Niepelt2024}. The second cost parameter, $\varphi$, is set to $1.503$, which is the midpoint of a range of values calculated by \textcite{Niepelt2024}. The fixed bank equity, $e=1.38$, is calibrated so that the share of total capital held by the bank is $0.3$. 

On the firm's side, we set the capital share of output, $\alpha$, and the capital depreciation rate, $\delta$, to the standard values of $1/3$ and $0.025$, respectively. The probability of not being able to adjust the retail price is $\theta=0.75$. We set the steady-state level of the inverse elasticity of substitution between good varieties, $\bar{\eta}$, to $1/11$. The investment adjustment cost, $\theta_c$, is set to 10. The steady state value of productivity, $\bar{a}$, is set to 1. The persistence parameters of the AR(1) shocks for $a_t$ and $\eta_t$ are set to $0.9$

\subsection{Government}
We assume that government spending in the steady state is $20\%$ of output. This implies a steady-state government spending of $\bar{g}=0.16$. We set the cost of issuing reserves to $\mu_r=0.0003$, in line with \textcite{Niepelt2024}. We assume that the cost of issuing CBDC equals the total resource cost associated with the bank's issuance of deposits, i.e., $\mu_m = (\omega+\zeta\mu_r)\lambda=0.002$. For conventional monetary policy, the Taylor rule parameters are standard. Monetary policy inertia is set to $\rho_R=0.5$. The policy responses to inflation and output are set to $\theta_\pi=1.5$ and $\theta_y=0.2$, respectively. In the baseline, the parameters of the CBDC Taylor rule follow those of conventional monetary policy (i.e., $\rho_R^m=0.5$, $\theta_\pi^m=1.5$, and $\theta_y^m=0.2$), such that the CBDC rate follows the policy rate with a constant spread. The government is assumed to target zero inflation in the steady state, i.e., $\bar{\pi}=1$. We target the reserve spread as $0.00497$, following \textcite{Niepelt2024}. This implies a steady-state reserve rate of $\bar{R}^r=1.005$. Lastly, the persistence of the government spending process is set to $0.9$.

\begin{table}[!htbp]
\centering
\caption{Model parameters}
\vskip 0.5em
\resizebox{\textwidth}{!}{
    \begin{threeparttable}
    \begin{tabular}{l*{4}{c}}
    \toprule \toprule
    \multicolumn{1}{c}{Parameter}&\multicolumn{1}{c}{Value}&\multicolumn{1}{c}{Source/Motivation}& \multicolumn{1}{c}{Description}\\
    \midrule 
    \multicolumn{3}{l}{Households}\\
    $\beta$ & $0.99$ & Standard & Discount factor\\ 
    $\sigma$ & $1$ & Standard & Risk aversion \\
    $\psi$ & $4.55$ & \textcite{Piazzesi2022_money} & Inv. elast. of sub. $c$ and $z$ \\
    $\iota$ & $1.0$ & \textcite{Chetty2011} & Inverse Frisch elasticity\\
    $v$ & $0.08$ & $z/y=1.04$ [\textcite{bayer}] & Liquidity utility weight \\
    $\xi$ & $9.46$ & $l=1/3$ & Labor disutility\\
    $\bar{\lambda}$ & $1$ & Assumption & Benefit of CBDC\\
    $\bar{\epsilon}$ & $1/6$ & \textcite{Bacchetta2022} & Inv. elast. of sub. $m$ and $n$\\
    \midrule 
    \multicolumn{3}{l}{Banks}\\
    $\phi$ & $0.0008$ & $\zeta=0.1945$ [\textcite{Niepelt2024}] & Operating cost \\ 
    $\varphi$ & $1.503$ & \textcite{Niepelt2024} & Operating cost \\ 
    $e$ & $1.38$ & $k^b/k=0.3$ & Equity\\ 
    \midrule 
    \multicolumn{2}{l}{Firms}\\
    $\alpha$ & $1/3$ & Standard & Capital share of output\\
    $\delta$ & $0.025$ & Standard & Capital depreciation rate\\
    $\theta$ & $0.75$ & Standard & Probability of fixed price\\
    $\bar{\eta}$ & $1/11$ & Standard & Inv. elast. of sub. between goods\\
    $\theta_c$ & $10$ & Standard & Investment adjustment cost\\
    $\bar{a}$ & $1$ & Standard & Productivity\\
    \midrule 
    \multicolumn{3}{l}{Government}\\
    $\mu_r$ & $0.0003$ & \textcite{Niepelt2024} & Reserves cost\\
    $\mu_m$ & $0.002$ & $\mu_m=(\omega+\zeta\mu_r)\lambda$ & CBDC cost\\
    $\rho_R$ & $0.5$ & Standard & Bond rate smoothing \\
    $\theta_\pi$ & $1.5$ & Standard & Bond rate response to $\pi$ \\
    $\theta_y$ & $0.2$ & Standard & Bond rate response to $y$\\
    $\rho_R^m$ & $0.5$ & Assumption & CBDC rate smoothing \\
    $\theta_\pi^m$ & $1.5$ & Assumption & CBDC rate response to $\pi$ \\
    $\theta_y^m$ & $0.2$ & Assumption & CBDC rate response to $y$\\
    $\bar{\pi}$ & $1$ & Standard & Inflation target \\
    $\bar{R}^r$ & $1.005$ & $\chi^r=0.00497$ [\textcite{Niepelt2024}] & Reserve rate\\
    $\bar{R}^m$ & $1$ & Assumption & CBDC rate \\
    \bottomrule
    \end{tabular}
\end{threeparttable}
}
\vskip 0.5em
\begin{minipage}{\textwidth}
    \footnotesize
    \textit{Note}: This table shows the baseline calibration. Persistence parameters for all AR(1) shocks are set to $0.9$.
\end{minipage}
\label{cal_params}
\end{table}

\section{Shock to the CBDC benefit}\label{s:IRFs}
One of the key questions when designing a CBDC concerns the public's perception of its usefulness. In this section, we examine how shocks to household preferences for CBDC are transmitted to the wider economy. Specifically, we consider a shock to the CBDC benefit, $\lambda_t$.

\begin{figure}[!htbp]
\centering
\caption{Impulse responses to a $25\%$ increase in CBDC benefit}
\begin{minipage}{\textwidth}
    \centering
    \makebox[\textwidth]{\includegraphics[width=1\textwidth]{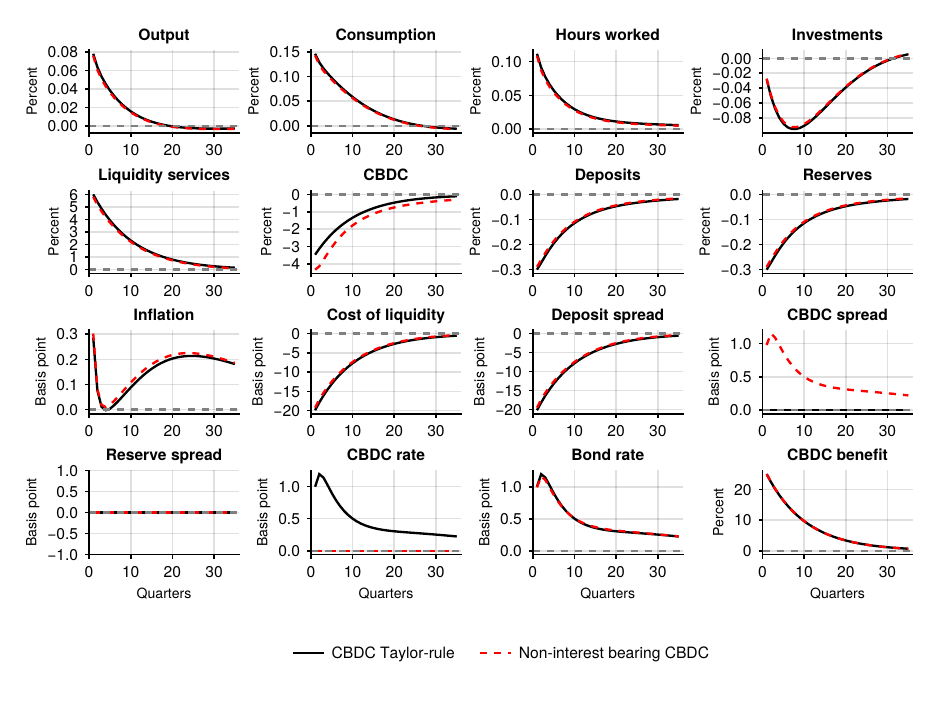}}
\end{minipage}
\vskip 1em 
\begin{minipage}{\textwidth}
    \footnotesize
    \textit{Note}: This figure shows the impulse responses of selected variables as percent deviation from steady state to a $25\%$ increase in the CBDC benefit, $\lambda_t$.
\end{minipage}
\label{f:irf_el}
\end{figure}

Figure \ref{f:irf_el} shows the impulse responses of selected variables to a $25\%$ increase in $\lambda_t$. In the baseline, we set the steady-state value $\bar{\lambda} =1$, such that CBDC is perceived as equally beneficial as deposits. The increase in $\lambda$ can be interpreted as a flight to safety toward CBDC during stress periods, reflecting its safer nature as a central bank liability, or as an increase in the liquidity benefits that CBDC provides as a means of payment. To analyze the effects of CBDC rate setting in the short run, we compare the impulse responses in our baseline specification, where the CBDC rate follows a Taylor-type rule as specified in equation (\ref{rm_taylor}), to one in which the CBDC is non-interest bearing, i.e., $R_t^m=1, \forall t$.

The immediate impact of the shock resembles a small but positive demand shock. Output, consumption, hours worked, and inflation all increase, while investments decline. The bond rate, which is the conventional monetary policy tool, increases in response. The interest spread on reserves remains unchanged, as the government sets the reserve rate to follow the policy rate proportionally according to equation (\ref{Rr}).

For the banking sector, the constant reserve spread implies that the bank's desired liquidity ratio, given by equation (\ref{zeta2}), remains unchanged. Consequently, the unit cost of issuing deposits through the bank's cost function equation (\ref{bank_cost}) is also unchanged. Thus, because the government sets the reserve rate such that it follows the policy rate one-for-one, there is no impulse through the bank's marginal cost that would incentivize the bank to change its deposit spread.

However, Figure \ref{f:irf_el} shows a substantial decrease in the deposit spread. This result arises due to the interplay between bank market power and CBDC competition. Recall that in the model, banks are regional monopolists in local deposit markets, and the equilibrium deposit spread, pinned down by the pricing equation (\ref{xn}), includes a marginal cost term and a markup term. The markup depends on the weighted average of two substitution elasticities: the elasticity of substitution between consumption and deposits, $1/\psi$, and the elasticity of substitution between CBDC and deposits, $1/\epsilon_t$. The variable $s_t$, given by equation (\ref{s}), governs the relative weight of these elasticities. When $\lambda_t$ increases, $s_t$ rises as well, placing more weight on the elasticity of substitution between CBDC and deposits. This reflects heightened competition from CBDC, making household demand for deposits more elastic and reducing bank market power. The resulting reduction in the deposit markup leads to a narrowing of the deposit spread. 

On the household side, the decrease in the deposit spread reduces the average cost of liquidity, incentivizing the household to hold more aggregate liquidity services. Interestingly, despite this increase in aggregate liquidity, Figure \ref{f:irf_el} shows that household holdings of both CBDC and deposits decline, albeit to different extents. This is because the higher $\lambda_t$ allows the household to derive greater liquidity services from each unit of CBDC they hold. A sufficiently large increase in $\lambda_t$ allows the household to economize on their liquid asset holdings. Then, the household pays less in terms of interest spreads for these assets while gaining more benefits from them. However, the reduction in CBDC holdings is much larger than that in deposits. This asymmetry arises because the CBDC spread does not decline under either specification, causing the relative price of CBDC to increase. Consequently, as shown in equation (\ref{mn}) and illustrated in Figure \ref{f:irf_el}, the ratio of CBDC to deposits decreases.

Consumption increases due to its complementarity with liquidity services. A lower cost of liquidity raises the household's current marginal utility of consumption. In other words, the opportunity cost of savings has increased, incentivizing the household to save less and consume more. At the same time, the household's marginal benefit of leisure is now higher than the marginal cost, leading the household to reduce labor supply. Finally, an increase in liquidity holdings leads to higher societal resource costs for liquidity provision, which, in turn, reduces capital accumulation and investments.

Overall, the impact of a $25\%$ surprise increase to the CBDC benefit on the economy is relatively small. The degree of bank disintermediation is low, as deposit outflows remain modest.

The differences across specifications are primarily reflected in the behavior of CBDC-related variables, as illustrated in Figure \ref{f:irf_el}. In the baseline scenario, where the CBDC rate follows a Taylor-type rule, the CBDC spread is kept constant since the parameters of the CBDC Taylor rule mirror those governing the bond Taylor rule. Consequently, the cost of liquidity, which is a weighted average of the deposit and CBDC spreads, is determined entirely by the decline in the deposit spread.

In contrast, when CBDC is non-interest-bearing (i.e., red, dashed responses in Figure \ref{f:irf_el}), the CBDC spread increases in tandem with the bond rate. However, this increase is not sufficient to offset the decline in the deposit spread. Again, the response of the deposit spread dominates, driving most of the change in the cost of liquidity. Notably, the quantity of CBDC decreases more significantly when it is non-interest-bearing, reflecting a larger increase in the CBDC spread.\footnote{
In our model economy, differences in the transmission of shocks across specifications are primarily driven by the strength of the conventional monetary policy reaction, i.e., through adjustments in the bond rate. Figure \ref{f:irf_ea} in Appendix \ref{sec:app:additional2} shows that following a $1\%$ increase in productivity, the bond rate declines by 30 basis points, leading to substantial differences in the CBDC spread and the cost of liquidity, depending on whether the CBDC rate is flexible or fixed. The difference in the cost of liquidity translates into different responses in economic allocation.}

\begin{figure}[!htbp]
\centering
\caption{Impulse responses to a $25\%$ increase in CBDC benefit with competitive banks}
\begin{minipage}{\textwidth}
    \centering
    \makebox[\textwidth]{\includegraphics[width=1\textwidth]{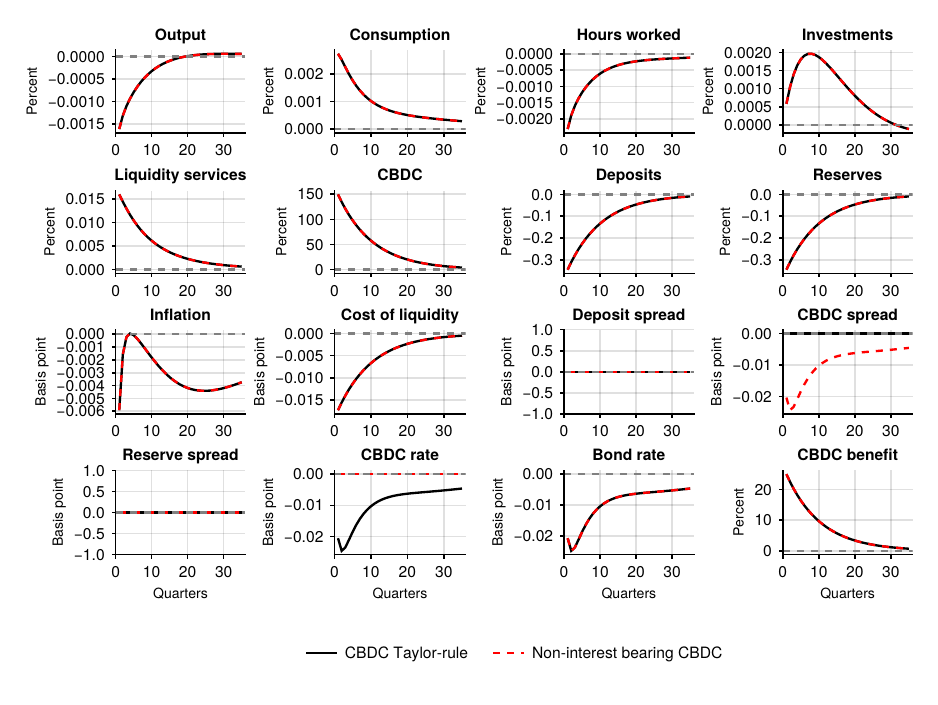}}
\end{minipage}
\vskip 1em 
\begin{minipage}{\textwidth}
    \footnotesize
        \textit{Note}: This figure shows the impulse responses of selected variables as percent deviation from steady state to a $25\%$ increase in the CBDC benefit to deposits, $\lambda_t$. In this specification, the banking sector is competitive.
\end{minipage}
\label{f:irf_cbank_el}
\end{figure}

Figure \ref{f:irf_cbank_el} illustrates the impulse responses to an increase in $\lambda_t$ under a competitive banking structure. With competitive banks, the markup term in the deposit spread equation (\ref{xn}) disappears, and the deposit spread is determined solely by the reserve spread and, indirectly, the bond rate. Since the reserve spread is constant, the deposit spread also remains constant. Without market power, the bank cannot adjust deposit pricing in response to increased competition from CBDC. As a result, the fall in the average cost of liquidity is much smaller compared to the case with non-competitive banks, and aggregate liquidity increases only marginally. This leads to the overall minimal response of the broader economy.

The large increase in CBDC is not as substantial as it may appear, since the steady-state quantity of CBDC is modest. Consequently, the percentage increase in CBDC remains small in absolute terms and should not be a concern, especially given the minimal change in total liquidity.

Considering expression (\ref{mn}) for the relative demand for CBDC, where the CBDC-to-deposit ratio depends on household preferences (the $\lambda$ and $\epsilon$ parameters) and on the relative costs of liquidity instruments, the similar degree of deposit disintermediation across banking structures suggests that the shift in preferences, rather than pricing dynamics, is the primary driver of the deposit response.\footnote{
To complement the analysis on the transmission of household preference shocks, Figure \ref{f:irf_ee} in Appendix \ref{sec:app:eps} shows the impulse responses to a $25\%$ increase in the inverse elasticity of substitution between CBDC and deposits, $\epsilon_t$, which reduces the substitutability between these two assets. Overall, the change in substitutability has a limited impact on economic allocation. Compared to an equal-sized shock to $\lambda_t$, the effects of changes in $\epsilon_t$ are significantly smaller for most variables.}

\section{Optimal interest rate rules for CBDC}\label{s:opt_mp}
Having analyzed the effects of CBDC preference shocks in the short run, we now turn to the welfare implications of CBDC interest rate policies. Specifically, we investigate how optimal Taylor-rule-based CBDC policies vary with the CBDC benefit and the structure of the banking sector.

To this end, we optimize simple policy rules, following the approaches of \textcite{Schmitt2007} and \textcite{Faia2007}, which offer a convenient framework for analyzing policy variables that can be described as linear functions of a few key model variables. In particular, we analyze the Taylor rule, as specified in equation (\ref{rm_taylor}), describing the nominal interest rate on CBDC. We do so by numerically computing the policy responses to inflation and output, $\theta_\pi^m$ and $\theta_y^m$ respectively, as well as the interest rate smoothing parameter, $\rho_R^m$, that maximize the expected value of welfare, $\mathcal{W}_t$, conditional on being in steady state. We rewrite the expected welfare in recursive form as
\begin{align}
    \mathcal{W}_t = \mathcal{U}(c_{t},z_{t+1},l_{t}) + \beta \mathbb{E}_t \mathcal{W}_{t+1},
\end{align}
and solve the model in the second order with pruning.

We limit the range of inflation and output responses to between $0$ and $4.0$ and the range of interest smoothing between $0$ and $0.99$. We follow \textcite{Schmitt2007} and set the standard deviations of the government spending shock, $g_t$, and productivity shock, $a_t$, to $0.016$ and $0.0064$, respectively. We set the standard deviation of the inverse elasticity of substitution between good varieties, $\eta_t$, to $0.14$, in line with \textcite{Ferrari2022}. We assume the standard deviations of shocks to the household preferences, $\lambda_t$ and $\epsilon_t$, to be $0.25$. Lastly, the standard deviations of the monetary policy shock and the CBDC Taylor rule shock are both set to $0$.

We consider household welfare and Taylor coefficients under six model specifications with different banking structures and CBDC benefit values. For each specification, we first calculate the optimal Taylor-rule coefficients for the standard policy rate and its maximized conditional welfare, assuming a non-interest-bearing CBDC (i.e., gross CBDC rate fixed at $1$). We then assess the welfare differences between a non-interest-bearing CBDC and CBDC Taylor rules, using the optimal coefficients for the bond rate from the initial step. In the baseline, we consider non-competitive banks and the benefit of CBDC equal to $\lambda=1$. Alternative specifications include competitive banks and cases where $\lambda=0.9$ or $\lambda=1.1$.

Table \ref{opt_mp} shows the welfare implications under different model specifications. Column (1) shows the reference conditional welfare under a fixed CBDC rate. Column (2) shows the welfare gains/losses, relative to the fixed rate case, given the baseline CBDC Taylor-rule coefficients (i.e., $\theta_\pi^m=1.5$, $\theta_y^m=0.2$, and $\rho_R^m=0.5$). Column (3) shows the welfare gains/losses, relative to the fixed rate case, given the optimized CBDC Taylor-rule coefficients. We quantify the welfare differences in columns (2) and (3) by expressing the gains/losses as compensating fractions of steady-state consumption-liquidity bundles.\footnote{
Note that given the non-separability between consumption and liquidity in our model, welfare gains/losses are expressed in terms of steady-state consumption-liquidity bundles, rather than consumption alone.
} For example, a compensating fraction of $1\%$ means that it would take $1\%$ of the steady-state consumption-liquidity bundles to equate the level of household welfare in the fixed CBDC rate case to the level of welfare under a CBDC Taylor rule.

Table \ref{opt_mp} illustrates that across all model specifications, implementing a Taylor-rule-based CBDC generally improves household welfare. Column (2) shows that, compared to a non-interest-bearing CBDC, welfare gains with standard coefficients for the CBDC interest rate Taylor rule that reacts to the state of the economy are small but positive (except for $\lambda$ lower than 1), ranging from $0.01\%$ to $0.03\%$.

As expected, from Column (3) it is evident that using optimized CBDC Taylor-rule coefficients improves welfare considerably, with gains ranging from $2.25\%$ to $7.04\%$ depending on the banking structure and CBDC benefit. From a preliminary investigation, these sizable welfare gains appear to stem from the CBDC interest rate's ability to stabilize productivity and markup (cost-push) shocks, acting as a complementary instrument to the standard policy rate. However, given the interactions across shocks, further analysis would be needed to isolate the dominant mechanisms driving these sizable welfare improvements.

With both monopolist and competitive banks, the biggest welfare gains occur when the benefit of CBDC is larger than 1. Intuitively, when the household derives more benefit from CBDC, having an optimized CBDC Taylor rule is then also more beneficial.

\begin{table}[!htbp]
\centering
\caption{Household welfare}
\vskip 0.5em
\begin{threeparttable}
    \begin{tabular}{l*{4}{c}}
    \toprule \toprule
    \multicolumn{1}{c}{}&\multicolumn{1}{c}{\specialcell{Fixed rate\\(1)}}&\multicolumn{1}{c}{\specialcell{Baseline Taylor\\(2)}}&\multicolumn{1}{c}{\specialcell{Optimized Taylor\\(3)}}\\
    \midrule 
    \multicolumn{3}{l}{Monopolist banks}\\
    \quad Baseline $\lambda=1$ & -123.8327 & 0.02\% & 4.34\% \\
    \quad Low $\lambda=0.9$    & -123.5857 & 0\%    & 4.41\%  \\
    \quad High $\lambda=1.1$   & -124.0111 & 0.03\% & 5.38\%  \\
    \midrule 
    \multicolumn{3}{l}{Competitive banks}\\
    \quad Baseline $\lambda=1$ & -126.2246 & 0.01\% & 4.16\% \\
    \quad Low  $\lambda=0.9$   & -126.2499 & 0.01\% & 2.25\% \\
    \quad High  $\lambda=1.1$  & -126.1864 & 0.02\% & 7.04\% \\
    \bottomrule
    \end{tabular}
\end{threeparttable}

\vskip 0.5em
\begin{minipage}{\textwidth}
    \footnotesize
    \textit{Note}: This table shows the welfare gains/losses of varying the interest rate on CBDC under different model specifications. Column (1) shows the reference conditional welfare under a fixed gross CBDC rate equal to 1. Column (2) shows the welfare gains/losses under the baseline CBDC Taylor-rule coefficients (i.e., $\theta_\pi^m=1.5$, $\theta_y^m=0.2$, and $\rho_R^m=0.5$). Column (3) shows the welfare gains/losses given the optimized CBDC Taylor-rule coefficients. Welfare gains/losses in columns (2) and (3) are expressed as the compensating fractions of steady-state consumption-liquidity bundles. 
\end{minipage}
\label{opt_mp}
\end{table}

Table \ref{opt_mp2} shows the optimized CBDC Taylor-rule coefficients associated with each model specification.\footnote{
Table \ref{opt_bond} in Appendix \ref{sec:app:opttr} shows the optimized policy rate Taylor-rule coefficients calculated in the first step of the optimization.
} 
When banks have market power, it seems generally beneficial for the CBDC rate to react to both inflation and output (except for $\lambda$ greater than 1). The output response is of similar magnitudes across the three specifications. In the baseline, optimal policy calls for an aggressive response to inflation while responding to output by about half as much. The weight on inflation response is smaller when the benefit of CBDC is lower than 1. In a scenario where CBDC provides a higher benefit, i.e., $\lambda=1.1$, the optimal policy calls only for a response to output.

When the banks are competitive, we have the stark result that the CBDC rate does not need to react to inflation, but reacts maximally to output fluctuations, regardless of the benefit of CBDC. Here, note that the coefficients for the standard policy rule, reported in Table \ref{opt_bond} in Appendix \ref{sec:app:opttr}, are at their maximum for inflation and minimum for output, in line with the literature on simple and implementable monetary policy rules for the standard policy instrument, with the welfare-maximizing policy response to inflation is typically very high, while the output response tends to be muted [\textcite{Schmitt2007}, \textcite{Ferrari2022}]. This result suggests that with two separate policy rules, the central bank can use one to stabilize inflation and the other to focus on output.

Finally, Column (3) suggests that interest rate smoothing for the CBDC rate is generally undesirable. 

\begin{table}[!htbp]
\centering
\caption{Optimized CBDC Taylor-rule coefficients}
\vskip 0.5em
\begin{threeparttable}
    \begin{tabular}{l*{4}{c}}
    \toprule \toprule
    \multicolumn{1}{c}{}&\multicolumn{1}{c}{\specialcell{Inflation response $\theta_\pi^m$\\(1)}
    }&\multicolumn{1}{c}{\specialcell{Output response $\theta_y^m$\\(2)}}&\multicolumn{1}{c}{\specialcell{Inertia $\rho_R^m$\\(3)}}\\
    \midrule 
    \multicolumn{3}{l}{Monopolist banks}\\
    \quad Baseline $\lambda=1$ & 4.0   & 1.872 & 0.0  \\
    \quad Low $\lambda=0.9$    & 1.889 & 1.841 & 0.0  \\
    \quad High $\lambda=1.1$   & 0.0   & 1.947 & 0.0  \\
    \midrule 
    \multicolumn{3}{l}{Competitive banks}\\
    \quad Baseline $\lambda=1$ & 0.0   & 4.0   & 0.0  \\ 
    \quad Low $\lambda=0.9$    & 0.0   & 4.0   & 0.0  \\   
    \quad High $\lambda=1.1$   & 0.0   & 4.0   & 0.0  \\   
    \bottomrule
    \end{tabular}
\end{threeparttable}

\vskip 0.5em
\begin{minipage}{\textwidth}
    \footnotesize
    \textit{Note}: This table shows the coefficients of the CBDC Taylor rule which maximizes the conditional welfare in different model specifications. 
\end{minipage}
\label{opt_mp2}
\end{table}

In summary, our welfare analysis of CBDC interest rate policy suggests that a non-optimized Taylor-rule-based CBDC yields small welfare improvements relative to a non-interest-bearing CBDC, but the gains are considerable when the coefficients are optimized. When CBDC provides a higher benefit to the households (i.e., $\lambda>1$), the welfare improvements from having CBDC Taylor rules are larger. The optimized CBDC Taylor-rule coefficients vary with the banking structure. In a non-competitive banking sector, inflation response depends on the benefit of CBDC, while the magnitude of the output response is consistent across specifications. With competitive banks, we get the stark result that the CBDC rate should focus entirely on output. Lastly, our results suggest that interest rate smoothing for CBDC is not desirable.

\section{Conclusion}\label{s:the_end_and_the_death}
This paper examines the role of a CBDC as a policy instrument by studying the transmission of household preference shocks and the welfare implications of CBDC interest rate policies. We incorporate CBDC and banks into a standard New Keynesian model, where households make portfolio choices between CBDC and deposits, and liquid assets compete for household resources. We model banks as non-competitive to account for the importance of bank market power in the transmission and efficacy of monetary policy [see, e.g., \textcite{Drechsler2017} and \textcite{Wang2022}].

First, we find that a positive shock to the benefit of CBDC has a mildly expansionary effect on the economy. Due to bank market power and competition from CBDC, the deposit spread falls by a large margin. As households economize on liquid asset holdings, they reduce both CBDC and deposit balances. However, the degree of bank disintermediation is low, as deposit outflows remain modest. 

Second, we find that implementing a Taylor-rule-based CBDC interest rate policy leads to modest welfare improvements compared to an economy with a non-interest-bearing CBDC, but the gains are considerable when the coefficients are optimized. Welfare gains are higher when the CBDC provides a higher benefit. The optimal policy response varies across banking structures. With monopolist banks, the optimized interest rate response to inflation depends on the benefit of CBDC, while the magnitude of the output response is consistent across specifications. With competitive banks, we get the stark result that the CBDC rate should focus entirely on output. Lastly, our results suggest that interest rate smoothing for CBDC is not desirable.

Despite central banks currently investigating the issuance of a non-interest-bearing CBDC, our findings contribute to the ongoing debate by highlighting the potential macroeconomic implications if CBDC were to be used as a policy instrument. Understanding how different CBDC designs interact with bank market power, deposit pricing, and monetary policy transmission is crucial for assessing their broader economic effects.

\newpage

\nocite{*}
\printbibliography[
    heading=bibintoc,
    title=References,
    ]

\newpage
\appendix

\section{Additional figures} \label{sec:app:additional2}
\renewcommand{\thetable}{A.\arabic{table}}
\renewcommand{\thefigure}{A.\arabic{figure}}
\setcounter{table}{0}
\setcounter{figure}{0}

\begin{figure}[!htbp]
\centering
\caption{Impulse responses to a $1\%$ increase in productivity}
\begin{minipage}{\textwidth}
    \centering
    \makebox[\textwidth]{\includegraphics[width=1\textwidth]{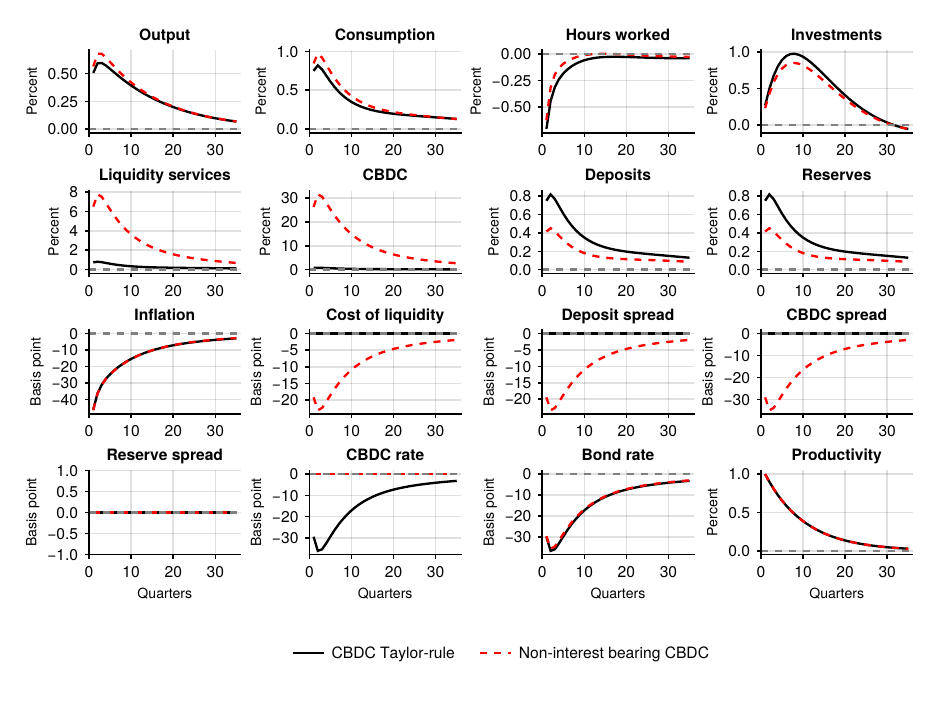}}
\end{minipage}
\vskip 0.5em 
\begin{minipage}{\textwidth}
    \footnotesize
    \textit{Note}: This figure shows the impulse responses of selected variables as percent deviation from steady state to a $1\%$ increase in productivity, $a_t$.
\end{minipage}
\label{f:irf_ea}
\end{figure}

\newpage

\section{Shock to the CBDC substitutability with deposits}\label{sec:app:eps}
\renewcommand{\thetable}{B.\arabic{table}}
\renewcommand{\thefigure}{B.\arabic{figure}}
\setcounter{table}{0}
\setcounter{figure}{0}
Figure \ref{f:irf_ee} illustrates the impulse responses to a $25\%$ increase in the inverse elasticity of substitution between CBDC and deposits, $\epsilon_t$, which reduces the substitutability between these two assets. Recall that to analyze the effects of CBDC rate setting in the short-run, we compare the impulse responses in our baseline specification, where the CBDC rate follows a Taylor-type rule as specified in equation (\ref{rm_taylor}), to one in which the CBDC is non-interest bearing, i.e., $R_t^m=1, \forall t$.

Despite the limited overall impact, a negative shock to the substitutability between assets appears mildly contractionary, except for a slight increase in consumption. The reserve spread remains unchanged (recall that the government sets the reserve rate to follow the policy rate proportionally according to equation (\ref{Rr})), while the deposit spread falls due to reductions in the deposit markup and the marginal cost of issuing deposits. Similar to a shock to the CBDC benefit, a higher $\epsilon_t$ increases the weight variable $s_t$, given by equation (\ref{s}), which affects the elasticity of demand for deposits. A lower substitutability between assets makes deposit demand more elastic, thereby reducing the deposit markup. The average cost of liquidity falls, and the household's holding of aggregate liquidity increases, which in turn contributes to a small rise in consumption.

A key difference between a substitutability shock and a shock to CBDC benefit is in the responses of CBDC and deposits. While household holdings of deposits still decrease, the decline is much larger in magnitude. Additionally, there is a clear substitution toward CBDC. In both cases, the deposit spread falls by more than the CBDC spread, making deposits a relatively cheaper source of liquidity. Based on equation (\ref{mn}), which pins down the CBDC-to-deposit ratio, this would induce an increase in deposit holdings relative to CBDC. However, the change in the relative price is accompanied here by the large increase in $\epsilon_t$. With the baseline calibration, the ratio of deposit spread to CBDC spread, $\chi_{t+1}^n / \chi_{t+1}^m$, falls below one after the shock. As a result, an increase in $\epsilon_t$ would increase the CBDC-to-deposits ratio. The substantial reduction in deposits is driven by the fact that the average cost of liquidity, $\chi_{t+1}^z$, decreases by a much larger margin than the deposit spread. From equation (\ref{n}), which governs household demand for deposits, it follows that deposits as a share of aggregate liquidity fall. The large increase in CBDC holdings reflects households compensating for the reduced substitutability between assets. As CBDC and deposits become less interchangeable, households require more CBDC to replace the reduced holdings of deposits.

\begin{figure}[!htbp]
\centering
\caption{Impulse responses to a $25\%$ increase in CBDC substitutability}
\begin{minipage}{\textwidth}
    \centering
    \makebox[\textwidth]{\includegraphics[width=1\textwidth]{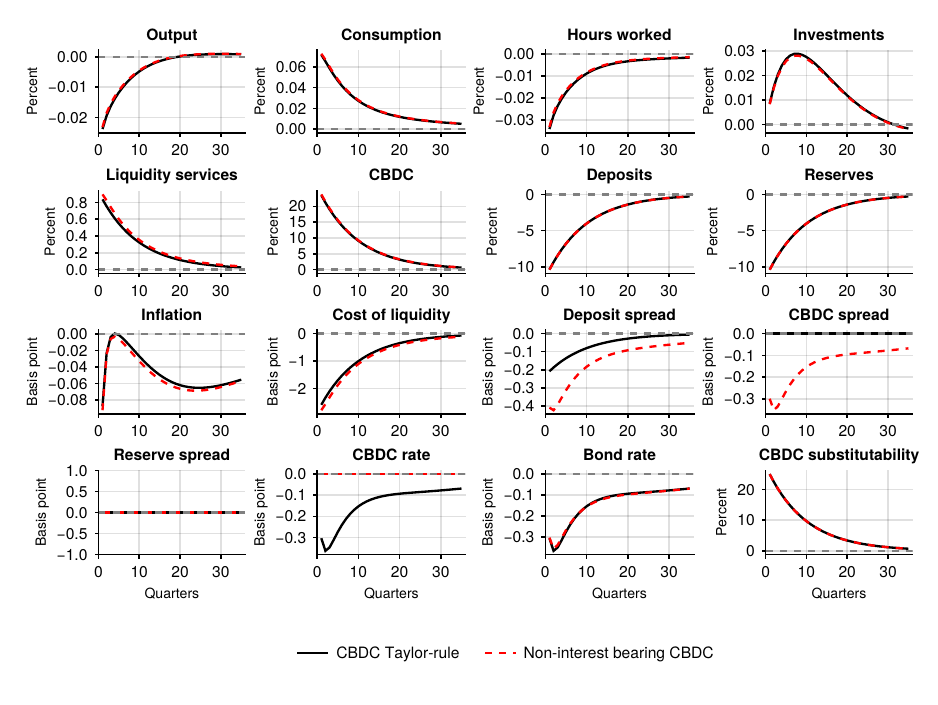}}
\end{minipage}
\vskip 0.5em
\begin{minipage}{\textwidth}
    \footnotesize
    \textit{Note}: This figure shows the impulse responses of selected variables as percent deviation from steady state to a $25\%$ increase in the inverse elasticity of substitution between CBDC and deposits, $\epsilon_t$.
\end{minipage}
\label{f:irf_ee}
\end{figure}

\begin{figure}[!htbp]
\centering
\caption{Impulse responses to a $25\%$ increase in CBDC substitutability with competitive banks}
\begin{minipage}{\textwidth}
    \centering
    \makebox[\textwidth]{\includegraphics[width=1\textwidth]{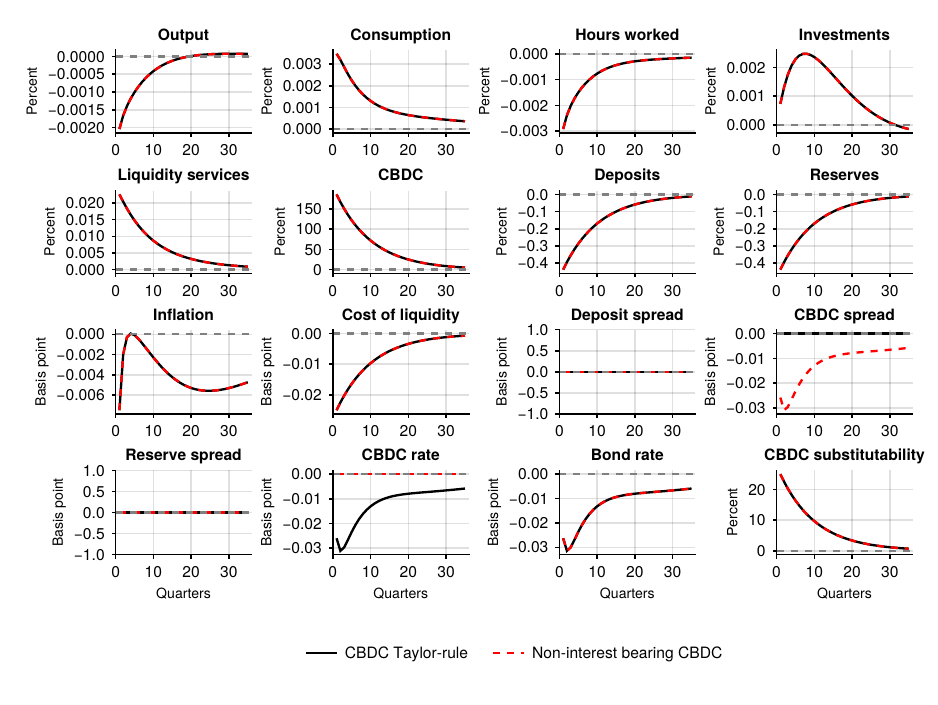}}
\end{minipage}
\vskip 0.5em
\begin{minipage}{\textwidth}
    \footnotesize
    \textit{Note}: This figure shows the impulse responses of selected variables as percent deviation from steady state to a $25\%$ increase in the inverse elasticity of substitution between CBDC and deposits, $\epsilon_t$. In this specification, the banking sector is competitive.
\end{minipage}
\label{f:irf_cbank_ee}
\end{figure}

Figure \ref{f:irf_cbank_ee} shows the impulse responses to $\epsilon_t$ when the banking sector is competitive. In this case, the already small effects of changes in $\epsilon_t$ are almost absent, underscoring the role of bank market power in amplifying the transmission mechanism.

\section{Optimal Taylor-rule coefficients} \label{sec:app:opttr}
\renewcommand{\thetable}{C.\arabic{table}}
\renewcommand{\thefigure}{C.\arabic{figure}}
\setcounter{table}{0}
\setcounter{figure}{0}

\begin{table}[!htbp]
\centering
\caption{Optimized bond Taylor-rule coefficients}
\begin{threeparttable}
    \begin{tabular}{l*{4}{c}}
    \toprule \toprule
    \multicolumn{1}{c}{}&\multicolumn{1}{c}{\specialcell{Inflation response $\theta_\pi$\\(1)}
    }&\multicolumn{1}{c}{\specialcell{Output response $\theta_y$\\(2)}}&\multicolumn{1}{c}{\specialcell{Inertia $\rho_R$\\(3)}}\\
    \midrule 
    \multicolumn{3}{l}{Monopolist banks}\\
    \quad Baseline $\lambda=1$ & 4.0 & 0.0 & 0.0  \\
    \quad Low $\lambda=0.9$    & 4.0 & 0.0 & 0.0  \\
    \quad High $\lambda=1.1$   & 4.0 & 0.0 & 0.0  \\
    \midrule 
    \multicolumn{3}{l}{Competitive banks}\\
    \quad Baseline $\lambda=1$ & 4.0 & 0.0 & 0.0  \\ 
    \quad Low $\lambda=0.9$    & 4.0 & 0.0 & 0.377  \\   
    \quad High $\lambda=1.1$   & 4.0 & 0.0 & 0.0  \\   
    \bottomrule
    \end{tabular}
\end{threeparttable}

\vskip 0.5em
\begin{minipage}{\textwidth}
    \footnotesize
    \textit{Note}: This table shows the coefficients of the bond Taylor rule which maximizes the conditional welfare in different model specifications. 
\end{minipage}
\label{opt_bond}
\end{table}


\end{document}